\documentclass[]{aa}
\usepackage{graphicx,natbib}
\usepackage{lipsum}
\usepackage{latexsym,amsmath,tabularx,amssymb}
\usepackage{mathrsfs}
\usepackage[latin1]{inputenc}
\usepackage[T1]{fontenc}
\bibpunct{(}{)}{;}{a}{}{,}
\usepackage[varg]{txfonts}

\def\mearth{M_{\oplus}}

\def\tdisc{T_{\rm disc}}

\def\f1{f_{\rm I}}
\def\msun{M_\odot}

\def\beq{\begin{equation}}
\def\eeq{\end{equation}}

\def\tdisc{T_{\rm disc}}

\def\mearth{M_\oplus}
\def\msun{M_\odot}

\def\mearth{M_\oplus}

\def\simgr{\,\hbox{\hbox{$ > $}\kern -0.8em \lower 1.0ex\hbox{$\sim$}}\,}
\def\simle{\,\hbox{\hbox{$ < $}\kern -0.8em \lower 1.0ex\hbox{$\sim$}}\,}
\def\beq{\begin{equation}}
\def\eeq{\end{equation}}

\def\simgr{\,\hbox{\hbox{$ > $}\kern -0.8em \lower 1.0ex\hbox{$\sim$}}\,}
\def\simle{\,\hbox{\hbox{$ < $}\kern -0.8em \lower 1.0ex\hbox{$\sim$}}\,}
\def\beq{\begin{equation}}
\def\eeq{\end{equation}}

\def\apj{ApJ}                 
\def\apjl{ApJ}                
\def\apjs{ApJS}               
\def\ao{Appl.Optics}          
\def\aap{A\&A}                

\def\({\left(}
\def\){\right)}
\def\<{\left<}
\def\>{\right>}

\def\({\left(} 
\def\){\right)} 
\def\<{\left<} 
\def\>{\right>} 

\def\bc{\begin{changebar}}
\def\bce{\begin{center}}
\def\beq{\begin{equation}} 

\def\bi{\begin{itemize}}

\def\btab{\begin{tabular}{p{1.7cm}p{12cm}p{1.5cm}}}
\def\bt2{\begin{tabular}{p{1 cm}p{4.5cm}p{10cm}}}

\def\ec{\end{changebar}}
\def\ece{\end{center}}
\def\eeq{\end{equation}} 
\def\ei{\end{itemize}}

\def\etab{\end{tabular}\\}

\def\mH2{m_\mathrm{H_2}}

\def\dmax0{\rho_\mathrm{max}}
\def\dmaxS0{\Sigma_\mathrm{max}}

\def\rH2{r_\mathrm{H_2}}

\def\r0max{r_\mathrm{0max}}

\def\s0{\sigma_\mathrm{0}}

\def\xp0{x_{\rm{M0}}}

\def\z0max{z_\mathrm{0max}}

\def\apj{{\it ApJ}}                 
\def\apjl{ApJ}                
\def\apjs{ApJS}               
\def\ao{Appl.Optics}          
\def\aap{{\it A\&A}}                
\usepackage{bm}
\usepackage{units}
\usepackage{color}
\usepackage{url}

\begin{document}

\title{A new metric to quantify the similarity between planetary systems - application to dimensionality reduction using T-SNE.}

\author{Y. Alibert \inst{1}}
\offprints{Y. Alibert}
\institute{Physikalisches Institut  \& NCCR PlanetS, Universit\"at Bern, CH-3012 Bern, Switzerland, 
        \email{yann.alibert@space.unibe.ch}}

\abstract
{Planet formation models are now often considering the formation of planetary systems, with more than one planet per system. This raises the question of how to represent planetary systems in a convenient way, (e.g. for visualisation purpose)  and how to define the similarity between two planetary systems, for example to compare models and observations.}
{We define a new metric to infer the similarity between two planetary systems, based on the properties of planets that belong to these systems. We then compare the similarity of planetary systems with the similarity of protoplanetary discs in which they form.}
{We first define a new metric based on mixture of gaussians, and then use this metric to apply a dimensionality reduction technique, in order to represent planetary systems (which should be 
represented in a high dimension space) in a 2 dimension space. This allows us study the structure of a population of planetary systems and its relation with the characteristics of protoplanetary discs in which planetary systems form.}
{We show that the new metric can help finding the underlying structure of populations of planetary systems. In addition, the similarity between planetary systems as we define in this paper is correlated with the similarity between the protoplanetary discs in which these systems form. {We finally compare the distribution of inter-system distances for a set of observed exoplanets, with the distributions obtained from two models: a population synthesis model, and a model where planetary systems are constructed by randomly picking synthetic planets. The observed distribution is  shown to be closer to the one derived from the population synthesis model than from the random systems.}}
{The new metric can be used in a variety of unsupervised machine learning techniques, like dimensionality reduction, clustering, etc., to understand results of simulations and compare them with the properties of observed planetary systems.} 

\keywords{planetary systems - planetary systems: formation - machine learning}

\authorrunning{Y. Alibert}
\titlerunning{T-SNE in planetary systems}

\maketitle

\section{Introduction}
\label{sec:introduction}

Since the discovery of the first exoplanet orbiting a solar type star (Mayor and Queloz 1995), numerous planets have been discovered, a non-negligible fraction of them being part of planetary systems, with more than one planet. One recent exemple is the discovery of the Trappist-1 system harbouring seven planets with know mass, radius, orbital elements and composition (e.g. Gillon et al. 2017, Grimm et al. 2018, Dorn et al. 2018). At the same time, different groups have developed theoretical and numerical models with the aim of computing the properties of planetary systems (Ida and Lin 2004, 2010, Alibert et al. 2013, Emsenhuber et al. in prep). One goal of these models is, by comparing theoretical results with observations, to improve our knowledge of the physical processes at work during planet formation.

These comparisons have in general focussed on the properties of planets (e.g. mass distribution, radius distribution, mass-radius correlation) but not on the global properties of planetary systems\footnote{one exception is the study of the distribution of period ratios in planetary systems, see e.g. Pfyffer et al. 2015}. Indeed, the dimensionality of a planetary system (the number of quantities that characterise it) scales with the number of planets in the system, and can be quite large. If one only considers, for example, mass and semi-major axis, one needs $2N$ parameters to characterise a system with $N$ planets. Comparing sets of data (e.g. results of simulations on one side, observations on the other side) is not easy as soon as the number of planet is larger than 1, and the purpose of the present paper is to propose a metric (or distance) in the space of planetary systems, in order to quantitatively compare different planetary systems. Such a distance can be used, for example, to discover classes of systems in simulations or observations. We emphasise the fact that throughout this paper, the word 'distance' will be used to designate the mathematical distance between two planetary systems in a high-dimension space, and  is not related with any physical distance (e.g. semi-major axis, physical distance between the observer and a star, etc...). In the same way, we will introduce later on the concept of 'density' of a planetary system, which is not related in any way to the physical bulk density of a planet.

The paper is organised as follows. We will in Sect. \ref{sec:distance} introduce the distance between planetary systems. We will then in Sect. \ref{sec:TSNE} use this distance to represent planetary systems (which belong intrinsically to a high dimension space) in a 2D space, using a dimensionality reduction technique named T-SNE. We will use the same technique to relate the similarity between planetary systems with the similarity of protoplanetary discs in which they form. Finally, we will discuss our results, in particular possible improvements of the distance we propose, in Sect. \ref{sec:conclusion}.

\section{Constructing a distance in the space of planetary systems}
\label{sec:distance}

\subsection{Properties of distances}
\label{subsec:distances}

We start by recalling the properties of a distance. A distance on a space $S$ is a function $d$ from $S^2$ to $\mathbb{R}^+$ with the following properties:
\begin{itemize}
\item $d(x,y) = 0 \Leftrightarrow x=y$
\item $d(x,y) = d(y,x)$
\item $d(x,y) \le d(x,z) + d(z,y)$ for every x,y,z in $S$
\end{itemize}
 A function $d$ which fulfils only the two first properties but not the last one (the triangular inequality) is called a pseudo-distance. It is important to remind that many machine learning techniques and algorithmes require the use of a distance and fail when using a pseudo-distance (Bishop, 2006, Goodfellow,  Bengio and Courville, 2016). For this reason it is important that the distance between planetary systems we will construct respects the triangular inequality.

In the special case of comparing planetary systems with only  one planet, an easy and natural way is to directly compute the Euclidian distance between the two points representing the planets in the feature space (e.g. mass and semi-major axis). For exemple, we can define the distance between the system $s_1$ and the system $s_2$ as:
$$
d_{\rm pla}(s_1,s_2) = \sqrt { (  (\log M_1 - \log M_2)^2 + (\log a_1 - \log a_2)^2 )}
$$
 where $M_1$, $M_2$, $a_1$, $a_2$ are the masses and semi-major axis of the planet in system 1 and system 2 respectively, expressed in some relevant unit. Note that we could also use directly $M$ and $a$ (and not their logarithm) to define the distance. Given the very large range of values these parameters can take, this would however be unpractical.   
This distance is the Euclidian distance, so it fulfils the three above mentioned properties. 

This definition gives the same importance to both the mass and semi-major axis. However, this can be changed by adding some positive constant factors $\alpha_M$ and $\alpha_a$ to obtain:
$$
d_{\rm pla}(s_1,s_2) = \sqrt { \alpha_M (  (\log M_1 - \log M_2)^2 + \alpha_a (\log a_1 - \log a_2)^2 )}
$$

For the rest of the paper, we will assume that planets are characterised only by two features: the logarithm of their mass $M$ and of their semi-major axis $a$. The generalisation to the case where more features characterise the planets (e.g. the radius, composition, etc...) is straightforward. 

\subsection{Density distance}
\label{subsec:raw_distance}

\begin{figure}
 \center
\includegraphics[height=0.3\textheight,width=0.3\textwidth,angle=-90]{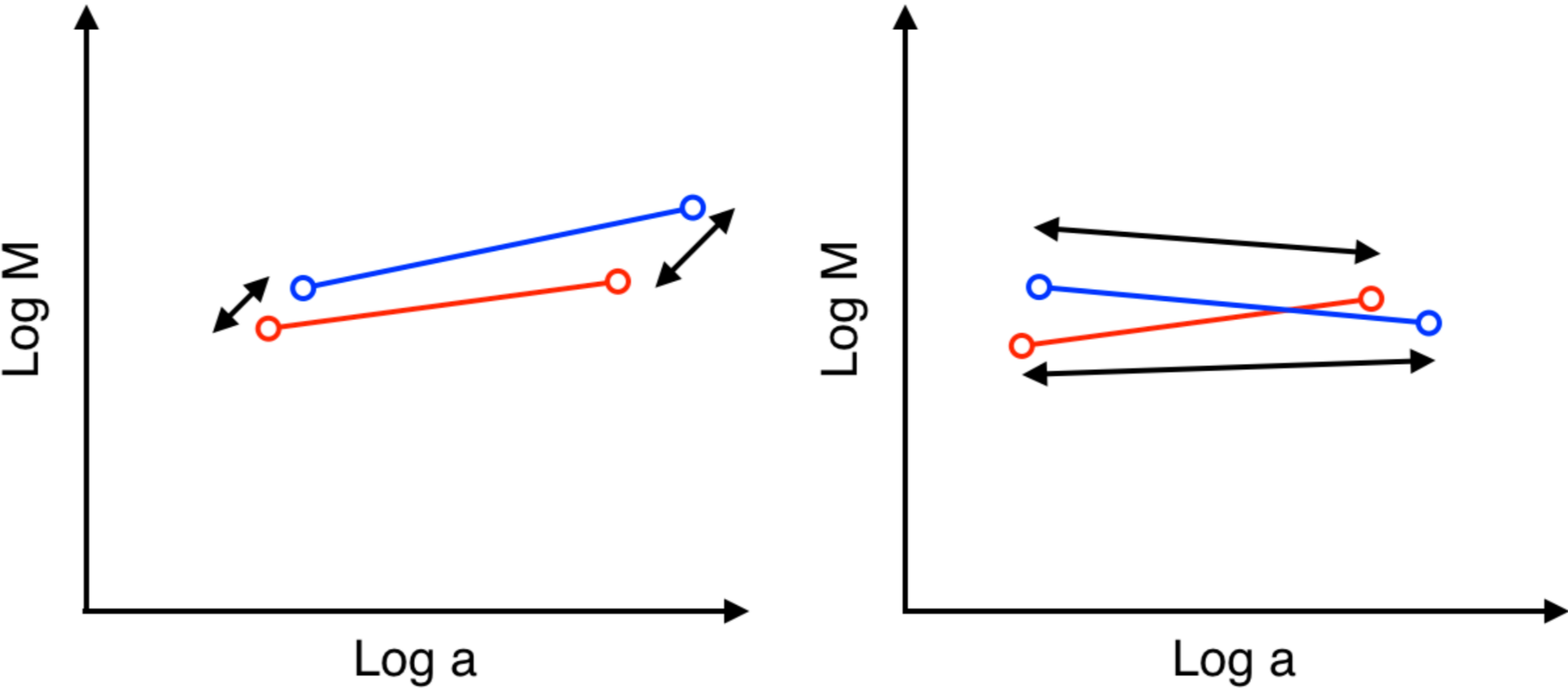}
\caption{Exemples of computation of distance between two planetary systems, one depicted in red, the other in blue. See text for details.}
  \label{ranking_mass}
\end{figure}

When more than one planet exist in each system, and assuming for the time that both systems harbour the same number of planets, we intuitively judge the similarity between two systems by looking how far each planet from $s_1$ is  from another planet from $s_2$. Still intuitively, we would like to define the distance between $s_1$ and $s_2$ as the sum (or the maximum) of these distances between planets (one belonging to $s_1$, the other belonging to $s_2$). Such an approach suffers however from the problem that there are many ways to construct such pairs of planets  (one in $s_1$, one in $s_2$) and the resulting distance between two systems may depend on the choice of these pairs. One way to avoid this arbitrariness is to rank the planets (for exemple by mass) in each system and to compare the lower mass planet in $s_1$ with the lower mass planet in $s_2$ and so on. However, this sill leads to some unwanted behaviour, as exemplified in Fig. \ref{ranking_mass}. On the left panel, we compare two systems (one depicted in red, the other in blue), and the ranking by mass leads to the comparison between the planets indicated by the arrows. The distance between these two systems is small. On the right panel, the system depicted in blue has been modified, the most massive planet being now the innermost one (and not the outermost one as on the left panel). In this case, the distance between the two systems is computed comparing the planets indicated by the arrows, and the resulting distance will be much larger than on the left panel. This is not satisfactory, since we intuitively would like to consider that the two situations are very similar, so the inter-system distances in both cases should be very close. 

One way to avoid the problem of choosing pairs is to use the Hausdorff distance (Hausdorff, 1914) defined as:
$$
d(s_1,s_2) = \max  \{  \sup\limits_{p \in s_1} \inf\limits_{q \in s_2} d_{\rm pla}(p,q),  \sup\limits_{q \in s_2} \inf\limits_{p \in s_1} d_{\rm pla}(p,q)  \}  
$$
The Hausdorff distance by definition fulfils the three properties of mathematical distances, but it gives the same importance to all planets in the system (see next section).

Another way to construct the distance between systems, and also avoid the problem of choosing pairs of planets, is to first define the 'density' of a system. For this, we first define, for each planet $p$ in a system $s$ a function $f_p(M,a)$given by:
$$
f_p(M,a) = \exp \left( - \left( {\log M - \log M_p \over 2 \sigma_m} \right)^2 -  \left( {\log a - \log a_p \over 2 \sigma_a} \right)^2 \right)
$$
 where $M_p$ and $a_p$ are the mass and semi-major axis of the planet, and $\sigma_M$ and $\sigma_a$ are constants, taken equal to $1/0.3$ for the rest of the paper (we discuss below and in Sect. \ref{sec:conclusion} the influence of these values). The function $f_p(M,a)$ 'smears out' the planet $p$, the constants $\sigma_M$ and $\sigma_a$ giving the scale of this smearing out in the $\log M$ and $\log a$ directions.  
 
Finally, we can define the 'density' $\psi_s$ of a system $s$ as:
$$
\psi_s (M,a) = \sum\limits_{p \in s}   f_p(M,a)
$$
 
\begin{figure}
 \center
 \includegraphics[height=0.5\textheight,angle=0]{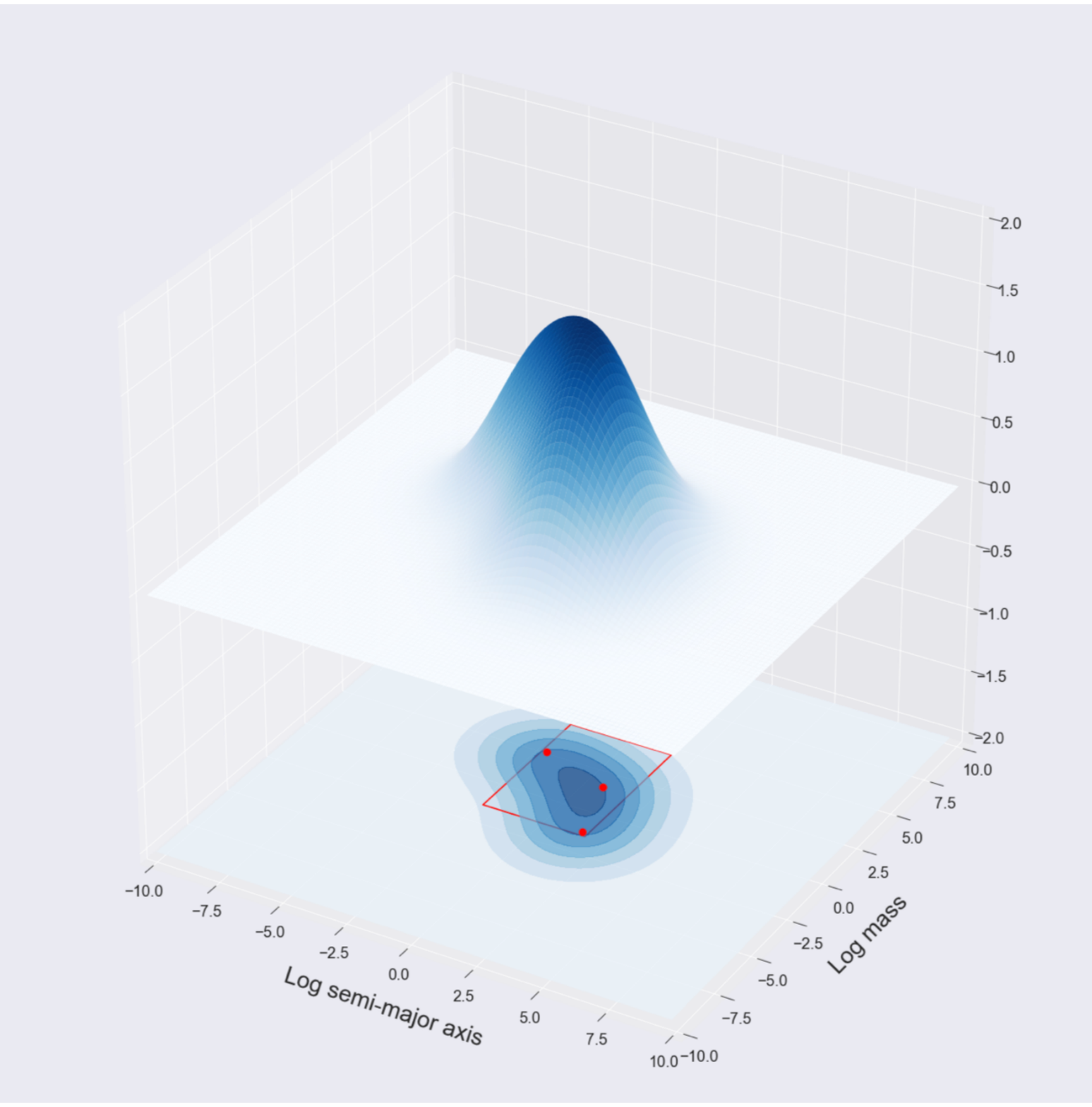}
\caption{Density of a planetary system $\psi$ with three planets (first planet of $100 \mearth$ at  $0.02 AU$, second of $10 \mearth$ at 1 AU, third of $0.02 \mearth$ at $\sim 80$ AU, represented as red points). The $\sigma_M$ and $\sigma_a$ parameters are both equal to $1/0.3$}
  \label{exemple_density}
\end{figure}

 Fig. \ref{exemple_density} gives an exemple of the density $\psi$ of a system with three planets. The red rectangle covers a range in $\log a$ extending from -2 to +2 (0.01 AU to 100 AU since $a$ is expressed in AU), and a range in $\log M$ extending from -2 to +4 (0.01 $\mearth$ to $10^4 \mearth$).
 
Finally, the distance between two systems $s_1$ and $s_2$ is given by:
$$
d(s_1,s_2) = \sqrt{ \int \left(\psi_{s_1} - \psi_{s_2}  \right)^2 d\log M d\log a }
$$
The integral should be extended to infinity, but in practical cases, it is sufficient to extend the integral to a large enough domain. In the exemples that follow, we integrate over a domain ranging from -10 to 10 both in $\log M$ and $\log a$.

The choice of the parameters $\alpha_M$ and $\alpha_a$ has some arbitrariness, and can be used to vary the importance of the mass and the semi-major axis in the determination of the distance. A small value of the parameter corresponds to a reduced importance of the corresponding quantity in the computation of the distance. In the limiting case of a value equal to 0, the corresponding parameter does not enter in the computation of the distance. Since these parameters correspond to the scale over which the planets are smeared out, a natural choice could be to use the observational uncertainties (in case some observations are considered).

\subsection{Weighted distance}
\label{subsec:weighted_distance}

One problem with the distances presented in the previous section is that all planets in a system have the same importance. Indeed, the functions $f_p$ all have the same integral, so a super-Jupiter and a sub-moon contribute in a similar way to the distance between two systems. One consequence is that the distance between a system $s_1$ and a system $s_2$ that is similar but contains in addition a tiny planet is non zero. In order to mitigate this effect, we introduce in the definition of the functions $f_p$ a weight that depends on the properties of the planet. Here again, the choice of the weighting is arbitrary, but must be the same for all systems. One possibility is to have a weight proportional to the logarithm of the mass of the planet, or proportional to the inverse of the period of the planet (planets located far from their star contributing less) or to its radial velocity semi-amplitude if the aim is to compare systems that are observed by radial velocity. In what follows, we choose to weight the function proportionally to the logarithm of the mass of the planet, independently of the period. In addition, the integral of the $\Psi$ function for each system is proportional to the logarithm of the total mass in the system.

\subsection{Distance distribution in a population of synthetic planetary systems}
\label{subsec:population}

The population of systems that we will use to illustrate the use of the distance presented in this paper have been computed using an updated version of the code of Alibert et al. (2005), Mordasini et al. (2009a,b), Mordasini et al. (2012a,b), Fortier et al. (2013), Alibert et al. (2013), Mordasini et al. (2015). In this model, we follow the growth and orbital evolution of ten planetary embryos in a protoplanetary disc, taking into account growth by gas and solid accretion, and orbital evolution by disc-planet interactions as well as planet-planet interactions. We do not take into account in these models enrichment of planetary envelopes by heavy elements (Venturini et al. 2015., 2016, Venturini \& Helled, 2017). The gas surface density in the initial protoplanetary disc  is given by:
$$
\Sigma = (2 - \gamma) { M_{\rm disc} \over 2 \pi a_C^{2-\gamma} r_0^\gamma } \left( {r \over r_0} \right)^\gamma \exp \left[ - \left( {r \over a_C}^{2-\gamma} \right) \right]
$$
where $r_0$ is equal to 5.2 AU, and $M_{\rm disc}$, $a_C$ and $\gamma$ are derived form the observations of Andrews et al. (2010). This gas surface density evolves as a result of viscous transport (in the framework of the $\alpha$ viscosity model) and photoevaporation (see references given above for details of the numerical model).
As in Mordasini et al. (2009a), the planetesimal-to-gas ratio is assumed to scale with the metallicity of the central star. For every protoplanetary disc we consider, we therefore select at random the metallicity of a star from a list of
$\sim 1000$ CORALIE targets (Santos, priv. comm.). Finally, following Mamajek et al. (2009), we assume that the cumulative distribution of disc lifetimes decays exponentially
with a characteristic time of 2.5 Myr. When a lifetime $\tdisc$ is selected (at random, following the afore-mentioned cumulative distribution), we adjust the photoevaporation rate in order that the protoplanetary disc mass reaches $10^{-5} \msun$ at the time $t = \tdisc$, when we stop the calculation. After the disc dispersal, the system is further evolved (computing planet-planet interactions and cooling of planets) for some time, the total simulated time (formation and evolution) being 20 Myr.
In each of these discs, ten planetary embryos are located at the beginning of the simulation. The initial location of the embryos is chosen at random, following a distribution uniform in logarithm.
The updated code is presented in Emsenhuber et al. (in prep), but we emphasise the fact that the focus of the present paper is to present the new metric. Application to the most recent simulations (with up to 50 or 100 planetary embryos growing in the same protoplanetary disc) will be presented in a future paper. 

\begin{figure}
\center
 \includegraphics[height=0.35\textheight,angle=0]{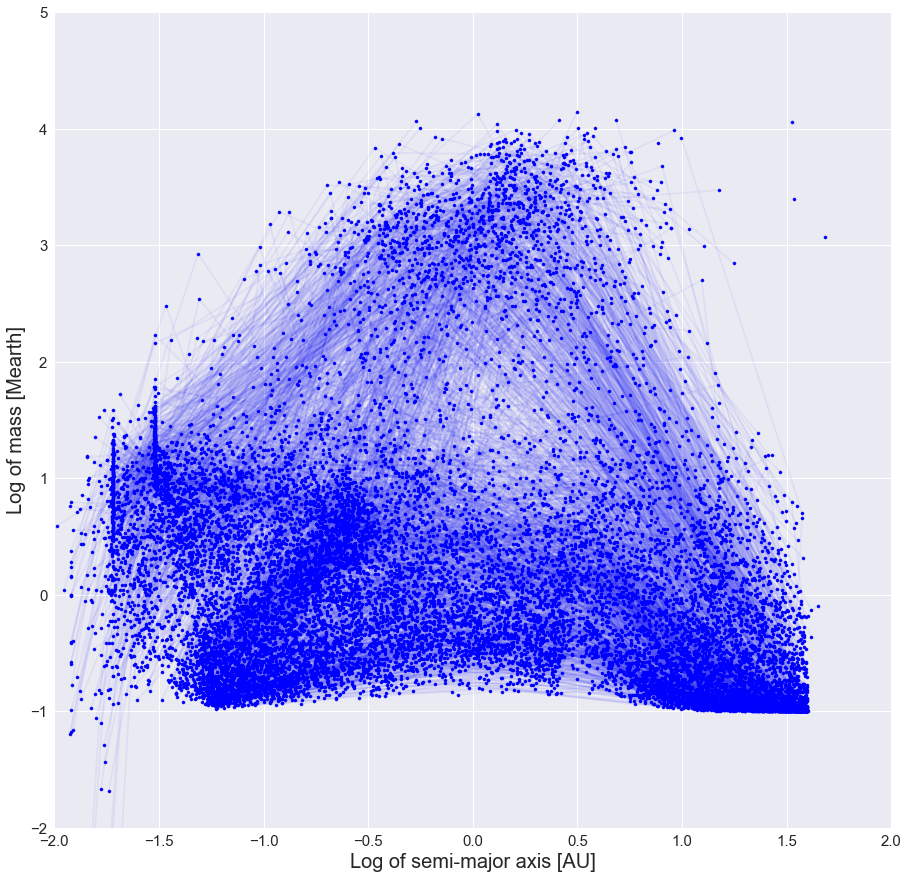}

  \includegraphics[height=0.35\textheight,angle=0]{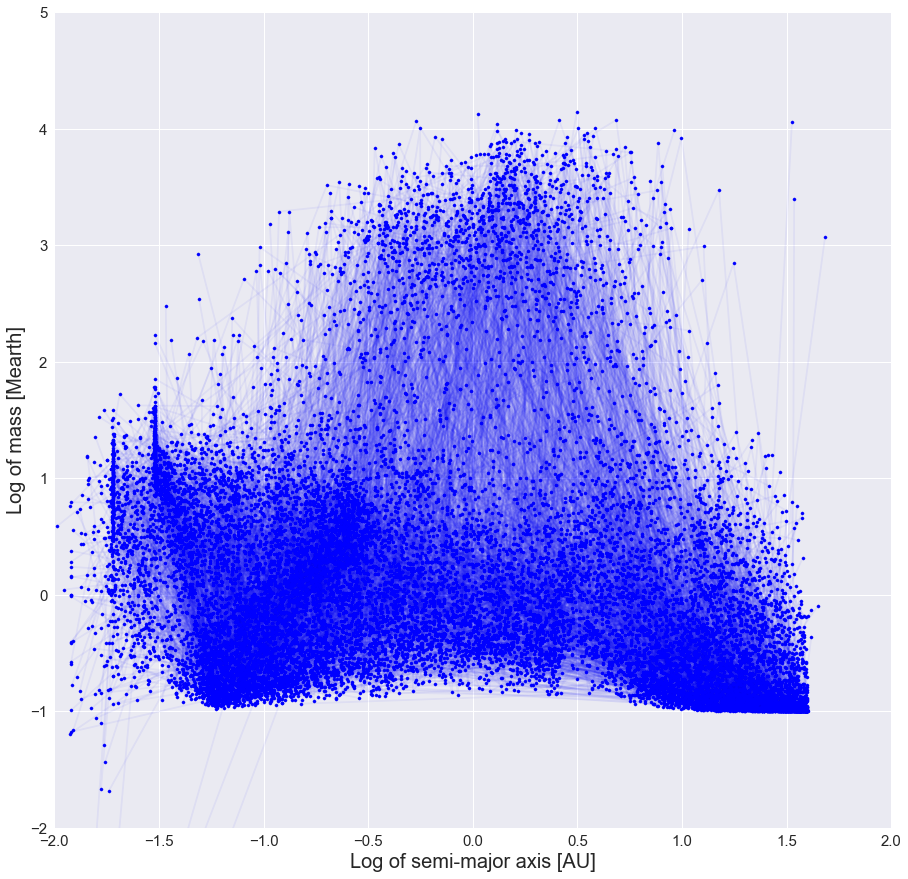}
\caption{Reference population (upper panel) and non-physical population (bottom panel). In each panel, planets are represented by points in the $\log a$-$\log M$ space (where $M$ is in $\mearth$ and $a$ is in AU). Planets belonging to the same system are linked by a line, a planetary system is therefore represent by a broken line with up to 9 changes of slope.}
  \label{ref_population}
\end{figure}

The population we use is shown in Fig. \ref{ref_population}, upper panel, where all planets belonging to the same planetary system are linked by a straight line. A planetary system is then represented as a broken line with up to 9 changes of slope, since it can contain up to 10 planets at the end of the simulation (some may be ejected, may collide with other planets or may be engulfed in the central star).

In order to compare the distribution of inter-systems distances with the one of another population, we constructed a set of non-physical systems in the following way. We took all the planets in our reference population  and produced new systems by drawing at random without replacement up to 10 planets. The distribution of the number of planets per system in these non-physical systems is the same as for the reference population. We show in Fig. \ref{ref_population}, lower panel, these non-physical planetary systems, where it can be seen already from the geometry of the broken lines that there are systematic differences between the two populations. We stress that we have not added any constraint during the construction of the  non-physical population, meaning that some of them could well be dynamically unstable or impossible to form.

\begin{figure}
 \center
 \includegraphics[height=0.35\textheight,angle=0]{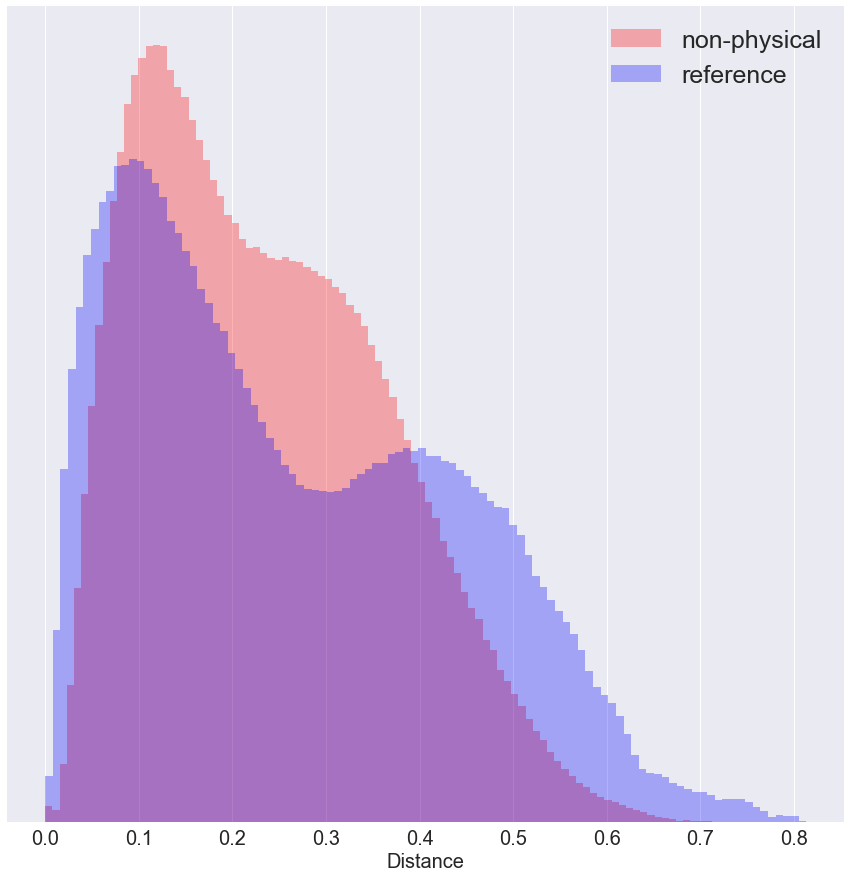}
\caption{Distribution of the inter-system distances in the reference population (blue) and the non-physical population (red).}
  \label{distribution}
\end{figure}

The distribution of the distances in the two populations is shown in Fig. \ref{distribution}, where it is clear that the non-physical systems are more similar one to each other (the distribution of distances - in red on the figure - is narrower). This is not surprising, since by shuffling planets between systems of the reference population, we have destroyed any correlation between planets in the same system, that could lead to systems being more dissimilar one to each other. This demonstrates the (expected) fact that a system of 10 planets produced by the numerical simulation is not just a collection of 10 independent planets.

\begin{figure*}
 \center
 \includegraphics[height=0.8\textwidth,angle=0]{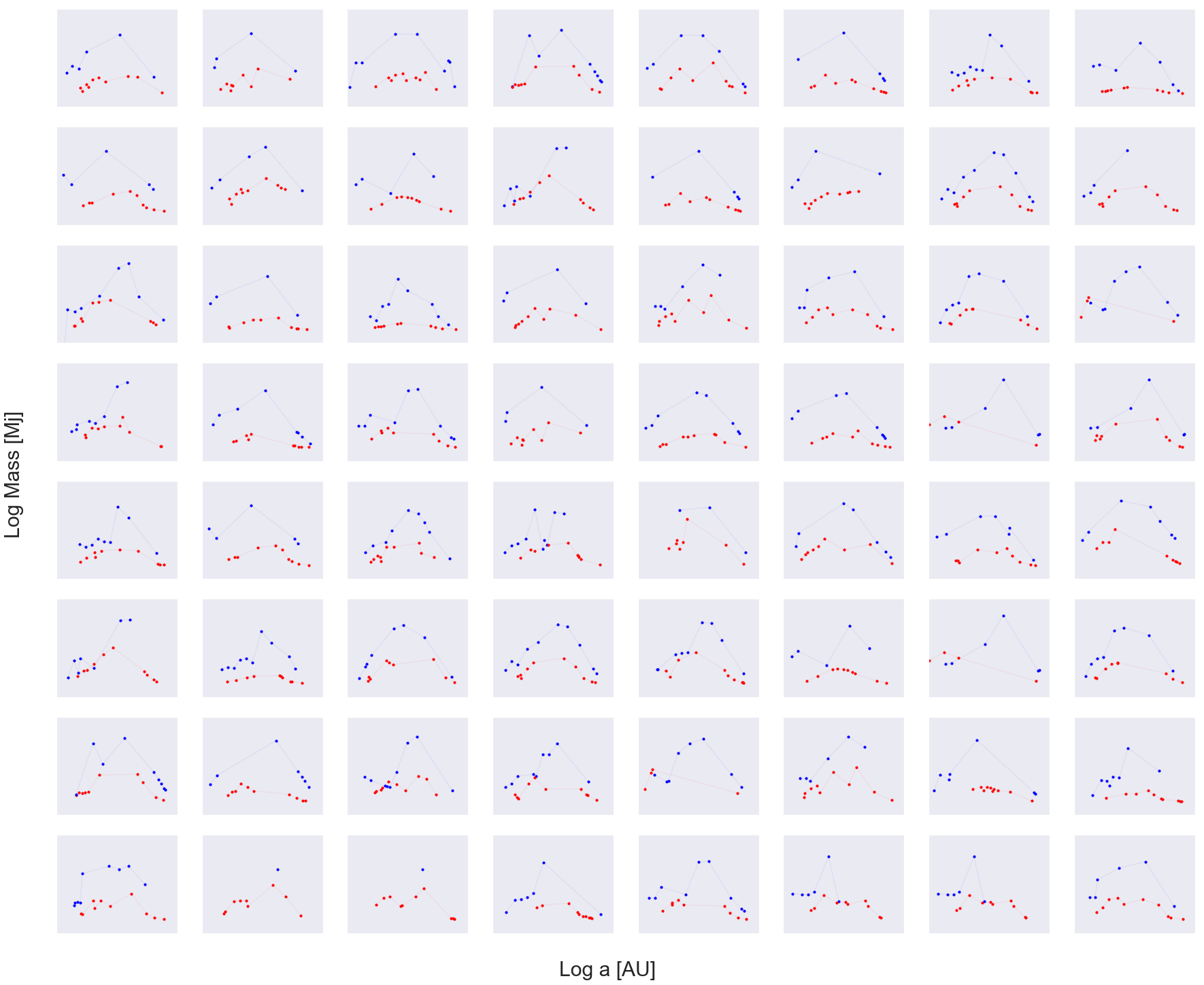}
\caption{Exemples of pairs of systems with mutual distance close to 0.4 (second peak in the distance distribution of Fig. \ref{distribution}) for the reference population. In each sub-panel, the system represented in red is the one with the smallest maximum mass. The axis in each panels have been omitted for clarity, the range for both axis is the same as in Fig. \ref{ref_population}.}
  \label{D04}
\end{figure*}

\begin{figure*}
 \center
 \includegraphics[height=0.8\textwidth,angle=0]{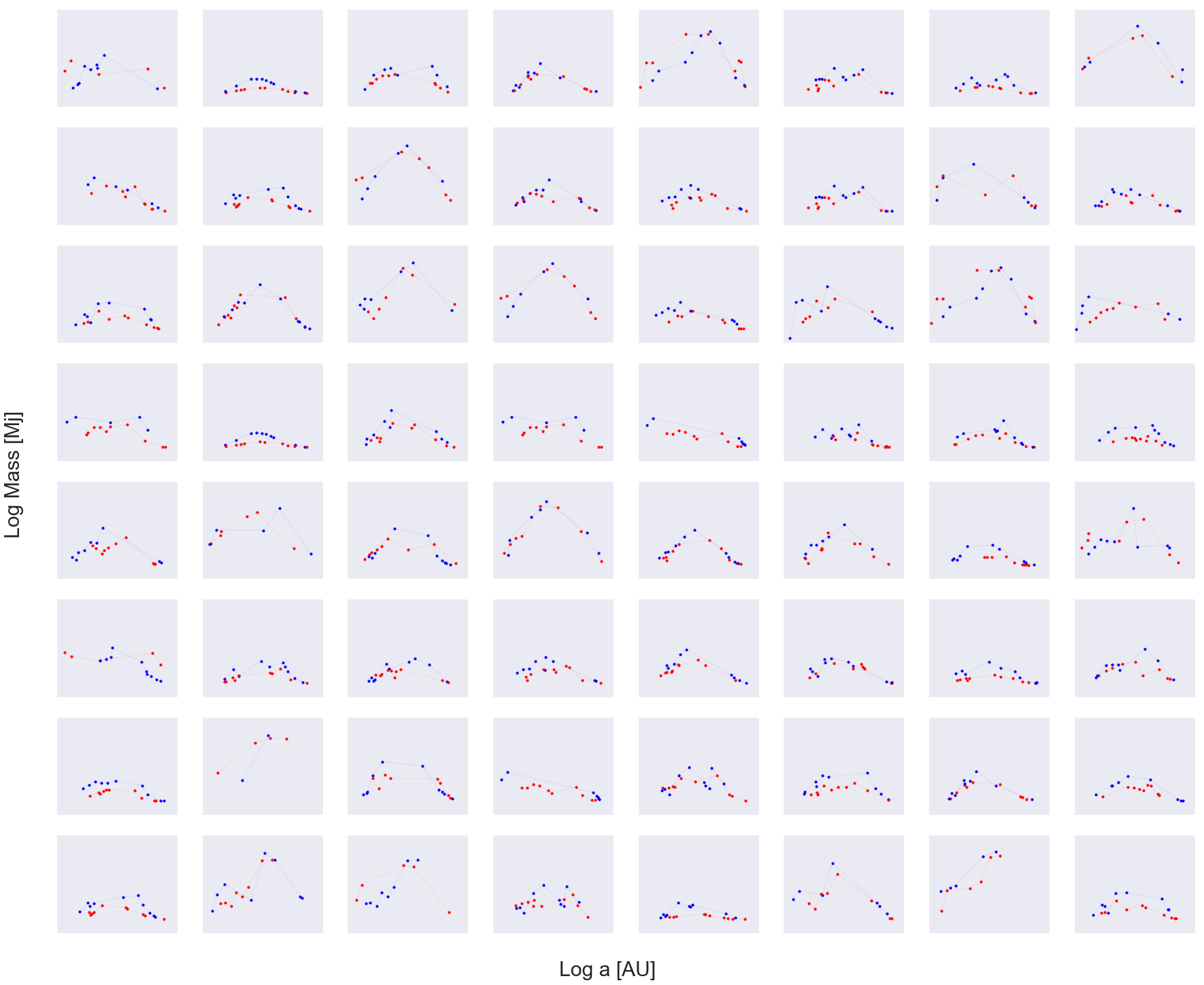}
\caption{Same as Fig. \ref{D04}, but for pairs of systems whose mutual distance is close to 0.1 (first peak in the distance distribution of Fig. \ref{distribution}).}
  \label{D01}
\end{figure*}

{The reference population shows two peaks, for distances around 0.1 and 0.4. We show in Fig. \ref{D04} and \ref{D01} pairs of systems with mutual distances equal to, respectively, 0.4 and 0.1. As can be seen on Fig. \ref{D04}, pairs of systems with mutual distance close to 0.4 (the second peak in the distance distribution) are  generally very dissimilar: one system harbours massive planets (in particular at intermediate semi-major axis), one system only harbours  small-mass planets. Moreover, the systems with massive planets (represented in blue) harbour in general fewer planets. The structure of the systems with small planets (represented in red) is generally  regular, and such systems are very unlikely to exist in the non-physical population. This explains the absence of a peak at similar large distance in the histogram shown in Fig. \ref{distribution} for the non-physical population. On the contrary, pairs of systems with mutual distance close to 0.1 shown in Fig. \ref{D01} are in general either both with small mass planets, both with larger planets, have similar numbers of planets, and are for some of them less regular than the ones shown in Fig. \ref{D04}.}

\begin{figure}
 \center
 \includegraphics[height=0.45\textwidth,angle=0]{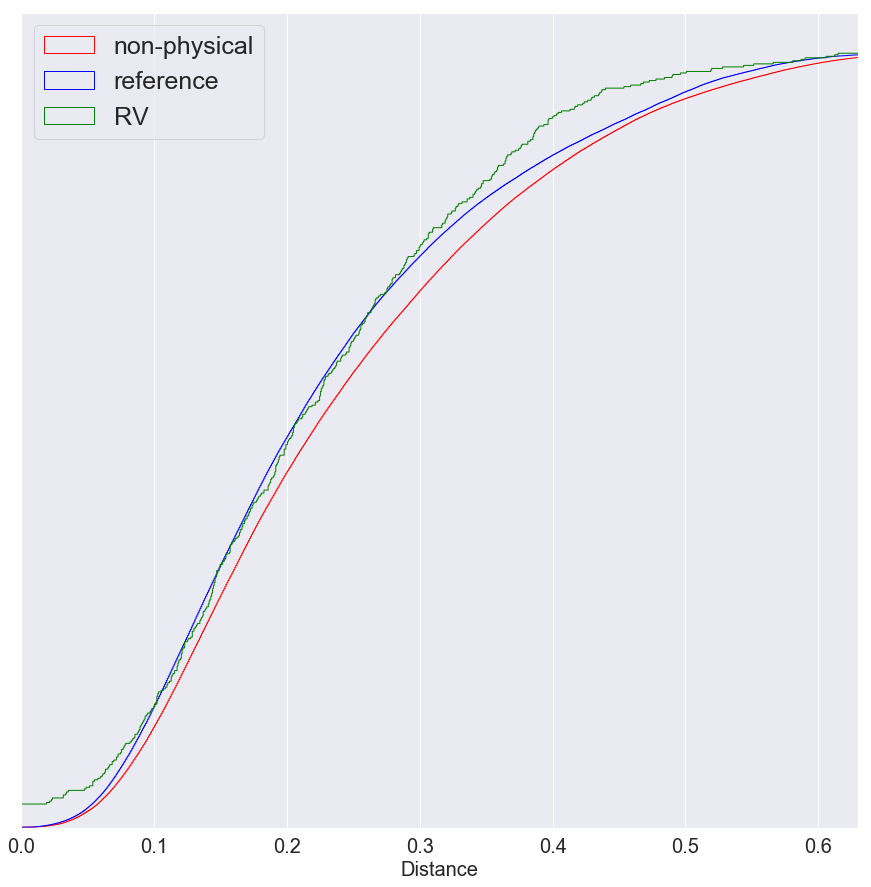}
\caption{Cumulative normalised distributions of the mutual distances for the reference, the non-physical and the RV populations. Only planets with period smaller than 5 years and radial velocity semi-amplitude larger than 3 m/s, and (for the RV population) orbiting around stars of mass comprised between 0.85 $\msun$ and 1.15 $\msun$ have been considered.}
  \label{compare_RV}
\end{figure}

{Although the populations used in this paper have not been computed with the most recent code (with up to 50 or 100 planetary embryos growing in the same protoplanetary disc - see Emsenhuber et al., in prep) we have compared in Fig. \ref{compare_RV} the cumulative distance distribution of both the reference and the non-physical populations, with a population of actual exoplanets planets detected (the 'RV' population). For this, we have selected planetary systems (with more than one planet) orbiting around stars of mass between 0.85 $\msun$ and 1.15 $\msun$ (our formation models assume a solar type central star). Since we use in this paper the mass and semi-major axis as primary planet parameters, we have only taken into account systems for which the masses of all known planets has been measured. We have then, for both synthetic populations (the reference and the non-physical ones), retained only planets whose period is smaller than 5 years, and radial velocity semi-amplitude is larger than 3 m/s. These values were chosen in order to approximately match the range of parameters of observed planets (RV population) we have considered\footnote{In order to be consistant, these cuts on the period and radial velocity semi-amplitude were also applied on the  planets of the RV population.}. The cumulative distribution of the RV population is closer to the one of the reference population than to the one of the non-physical population, although the match between the RV and the reference population is anyway not perfect.}

{One can finally note that the distributions of the two simulated populations are much closer one to the other than the cumulative distributions corresponding to Fig. \ref{distribution}. This results from considering only planets with small period and large radial velocity semi-amplitude.}

\section{Systems representation using low dimension embedding}
\label{sec:TSNE}

\subsection{T-SNE}

The distance we propose in the present paper can be used in the framework of unsupervised machine learning, for exemple for dimensionality reduction. Indeed, as already pointed out in the introduction, representing a planetary system with $N$ planets, each of them being characterised by two quantities, requires a space of dimension $2N$. The goal of dimensionality reduction algorithms is to represent the planetary systems in a space of dimension 2 (or 3) while keeping as much as possible the information regarding their repartition in the space of dimension $2N$.

Different dimensionality reduction algorithms have been developed, we will use here  T-SNE (for t-based stochastic neighbour embedding, see van der Maaten and Hinton, 2008) to represent systems of up to 10 planets in a 2 dimensions space. 
The T-SNE algorithm (van der Maaten, 2014) works in two steps. In a first step, the joint probability of two systems is computed. The joint probability between systems $i$ and $j$ depends on the distance between two systems (computed using the distance presented above) as:
$$
p_{i,j} \propto \exp \left( - {  d(s_i,s_j)^2 \over 2 \sigma^2} \right)
$$ 
where $\sigma$ is a parameter called the perplexity that controls the number of neighbours (see van der Maaten, 2014). 
Then, an iterative algorithm is used to minimise a cost function given by the Kullback-Leibler divergence (Kullback and Leibler, 1951) of the $p$ distribution and the $q$ distribution, where $q$ is the joint probability of two systems in the 2D space, function of the distance between the points representing systems $i$ and $j$ in the 2D space, and assumed to follow a t-student distribution with one degree of freedom (also known as Cauchy or Lorentz distribution):
$$
q_{i,j}  \propto \left( 1 + || y_i - y_j  ||^2  \right)^{-1}
$$
where $|| .  ||$ is the Euclidian norm in the 2D space.
The Kullback-Leibler divergence from $q$ to $p$ (also called relative entropy) is given by:
$$
D(p||q) = \sum\limits_{i,j \in S} p_{i,j} \log \left( { p_{i,j} \over q_{i,j} }  \right) 
$$
where $S$ is the population of systems we consider. This  function measures the loss of information occurring when using the $q$ distribution instead of the $p$ distribution. 
If two points are similar (large $p_{i,j}$ or small distance) in the $2N$ dimension space, they have to be close in the $2D$ space in order to avoid a large cost. If, on the other hand, two points are very dissimilar (large distance in the $20$ dimension space), there is no real constraints as the contribution to the cost function is anyway small, whatever the value of $q_{i,j}$.

An important point of the T-SNE algorithm is that the cost function is not convex (it is in particular invariant by translation and rotation in the 2D space). As a consequence, the result of T-SNE is non unique and it is in general advised to run a number of times the algorithm, changing slightly the initial position of the systems in the 2D space, in order to distinguish features that are robust from spurious structures.

\begin{figure}
\center
 \includegraphics[height=0.3\textheight,angle=0]{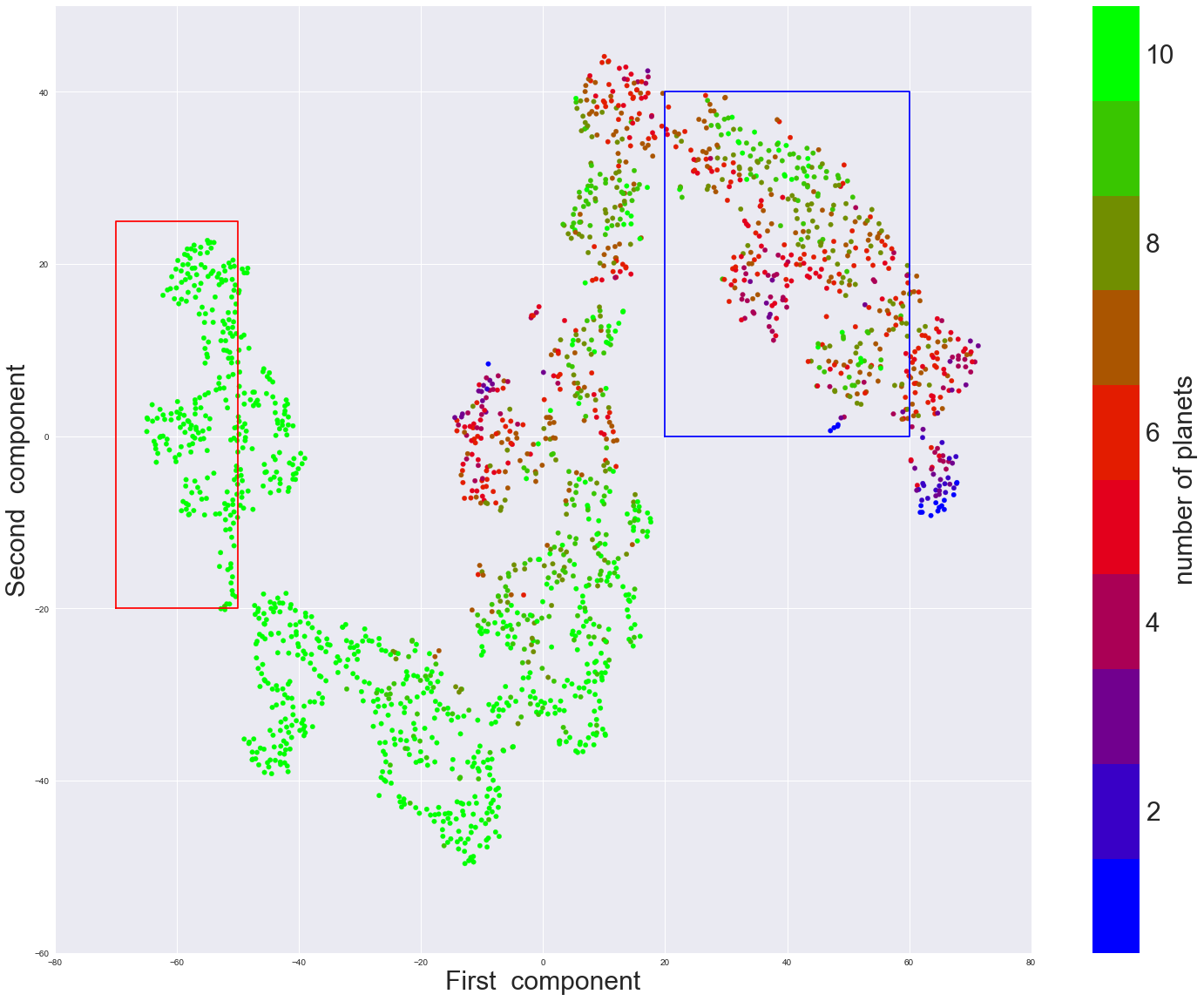}

   \includegraphics[height=0.3\textheight,angle=0]{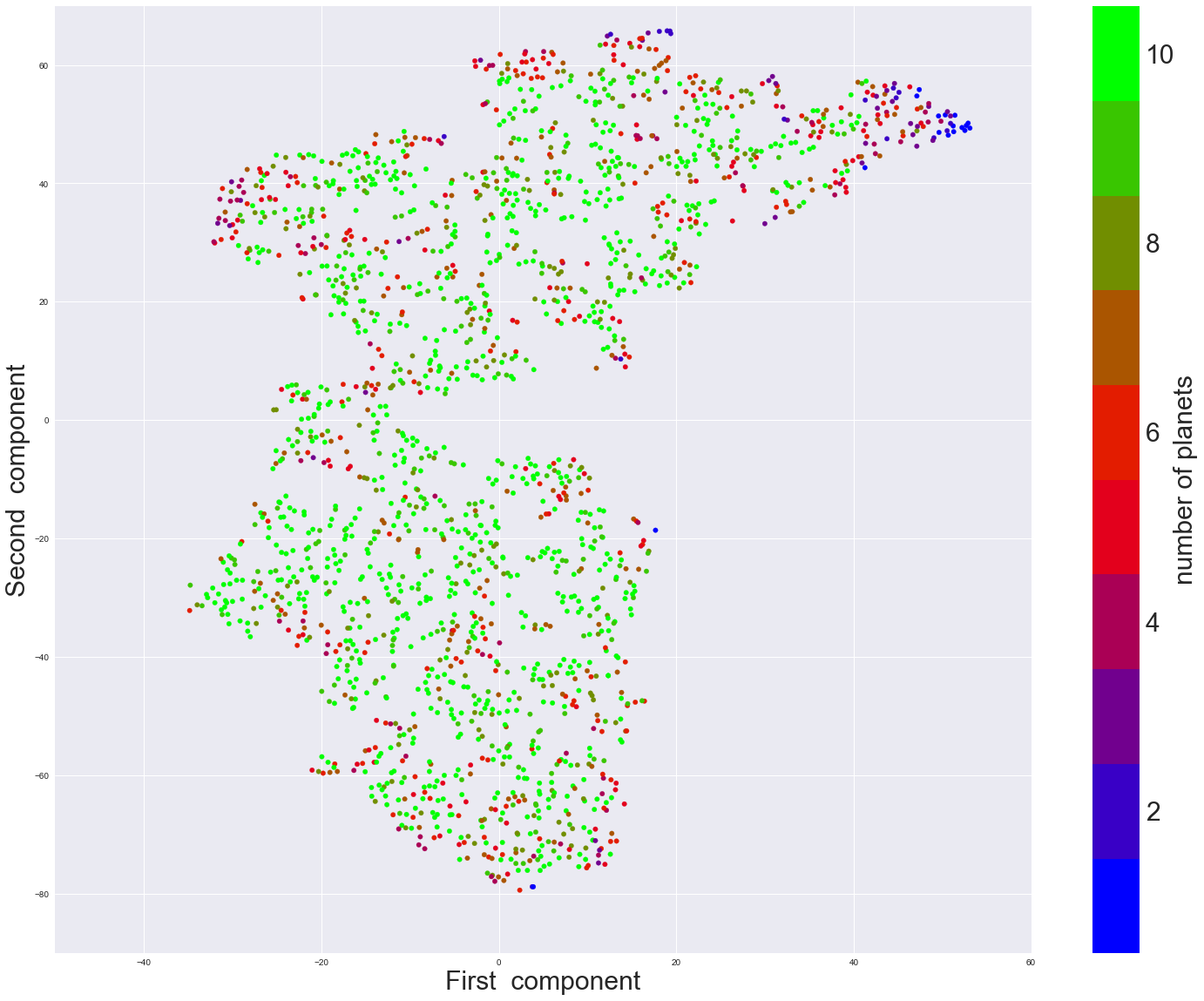}
\caption{T-SNE visualisation of the reference population (upper panel) and the non-physical population (lower panel). The colour code indicates the number of planets that remain at the end of the planetary system formation model. See text for the meaning of the blue and red rectangles.}
  \label{figTSNE}
\end{figure}

The result of the T-SNE visualisation for the two populations we consider is shown in Fig. \ref{figTSNE}, where the colour code indicates the number of planets at the end of the simulation. It is important to emphasise that a planetary system is represented in this diagram by a single point, as opposed to Fig. \ref{ref_population} where a planetary system is represented by a broken line with up to 9 changes of slope. In addition, the number of planets is \textit{not} an input of the T-SNE algorithm which only uses the mutual distance between systems. Finally, it is important to note that the T-SNE components of the systems have no physical meaning and cannot be related to the physical properties of the systems or the planets belonging to them.

\begin{figure}
\center
  \includegraphics[height=0.35\textheight,angle=0]{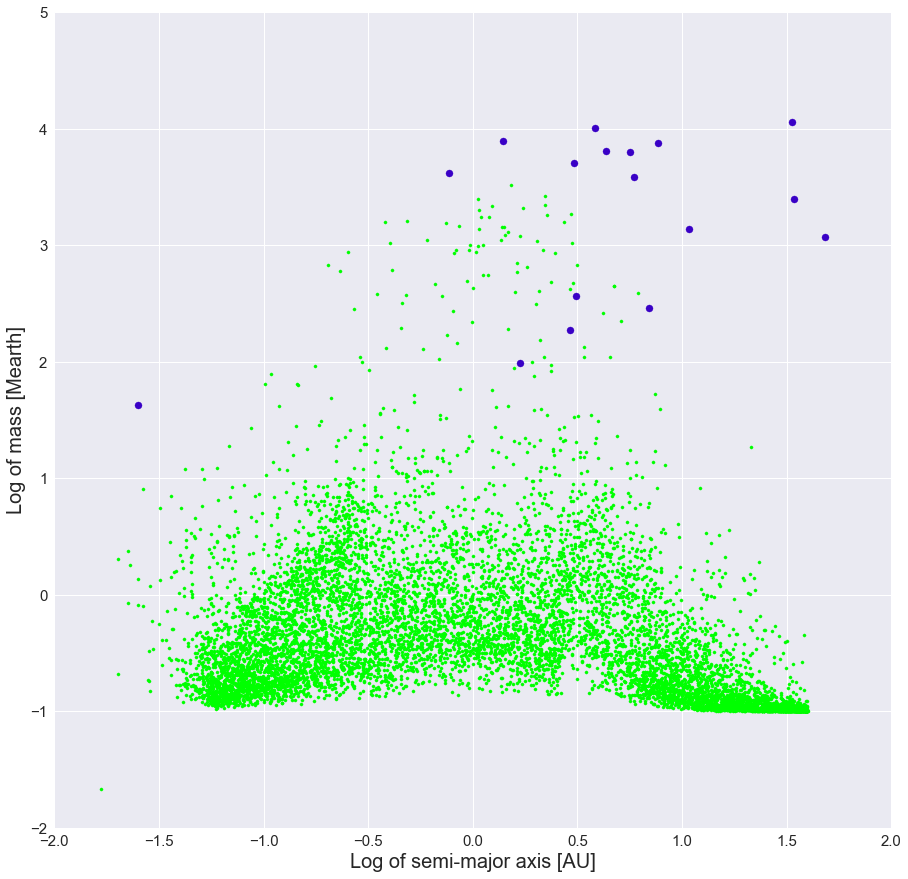}
\caption{Systems of the reference population with 10 planets (green) and one planet (blue).}
  \label{comp10-1}
\end{figure}

 On the upper panel, in the case of the reference population, a  non random distribution is seen, where systems  with the same (or nearly the same) number of planets lie close together. This means that these systems are also close (or similar) in the 20 dimension space. On the contrary, systems with only one planet lie far from all systems with 10 planets and they are therefore very different. This can be confirmed by examining these two classes of systems (10 planets versus 1 planet) in the $\log a$-$\log M$ space (see Fig. \ref{comp10-1}). As can be seen on this figure, systems with only one planet have a very different architecture (massive planet, located in general far from the central star) compared to 10-planets systems (in general small mass planets with a wide range of semi-major axis).
 
\begin{figure}
\center
  \includegraphics[height=0.35\textheight,angle=0]{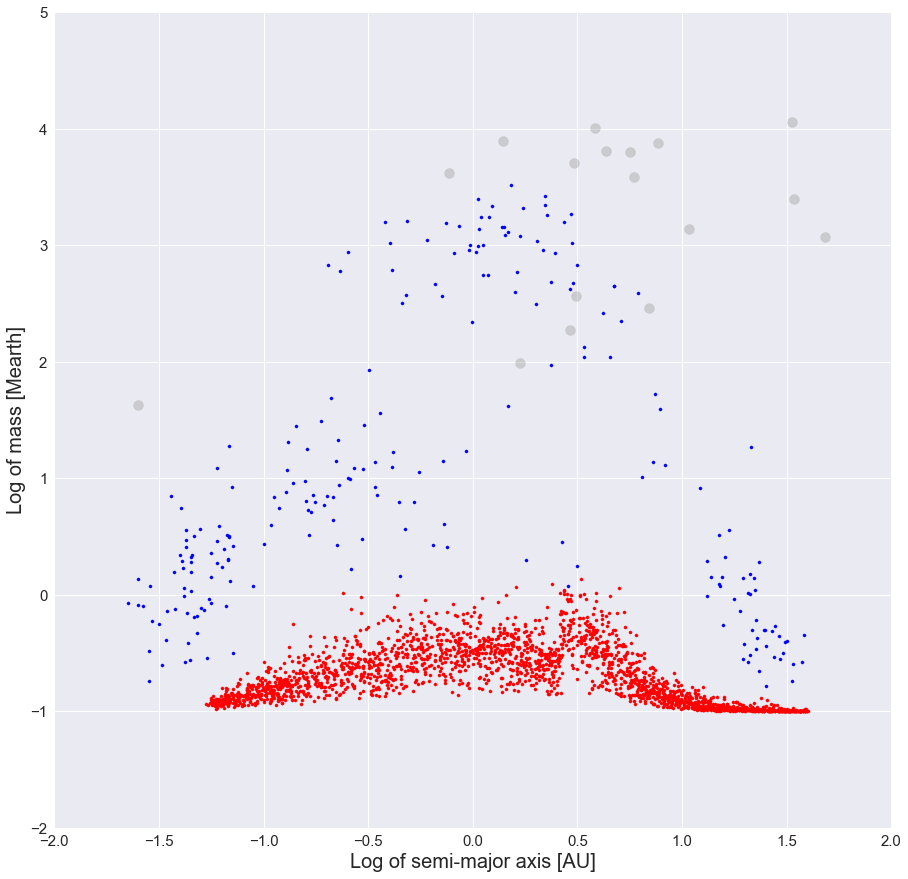}
\caption{Systems of the reference population with 10 planets whose  T-SNE representation lie in the red and blue rectangles in the upper panel of Fig. \ref{figTSNE}. The gray large dots are systems with only one planet, which are more similar to the systems represented in blue than to the one represented in red.}
  \label{comp10}
\end{figure}

  Another interesting feature is that systems with 10 planets (represented by light green points) are not clustered in the same part of the diagram. Comparing the 10-planets systems on the left part of the diagram (in the red rectangle) and on the right part (in the blue rectangle), in the  $\log a$-$\log M$ space (see Fig. \ref{comp10}), we see that these two classes correspond to systems with respectively only low-mass planets or only more massive planets. These two classes are very well separated, planets represented in red and in blue lying on both side of a part of the $\log a$-$\log M$ diagram where very few planets exist\footnote{The presence of this region with very few planets results from the fact that we have not considered 10-planets systems outside the blue and red rectangles, as can be seen by looking at Fig. \ref{comp10-1} which shows in light green \textit{all} planets belonging to 10-planets systems.}. Finally, the systems depicted in blue  in Fig. \ref{comp10} (in the blue rectangle of Fig. \ref{figTSNE}) are located closer to systems with only one planet on the T-SNE representation (Fig. \ref{figTSNE}). This is to be expected since the planets in these systems are more massive, and more similar to the planets in the 1-planet systems (Fig. \ref{comp10-1}).

In the case of the non-physical population (lower panel of Fig, \ref{figTSNE}), the distribution is very different with little spatial segregation between systems with a different number of planets. This again shows that there is a structure in the reference population that is lost when constructing the non-physical systems. 

\subsection{Link with protoplanetary disc properties}

An important question is wether the similarity between planetary systems is linked to the similarity of the protoplanetary discs in which they form. In order to answer this question, we need first to define the similarity between discs in the space of disc parameters. In our simulations, each disc model  depends on five parameters ($\Sigma_{\rm gas}$, $\Sigma_{\rm solids}$, $a_C$, $\gamma$, photoevaporation rate) where  $\Sigma_{\rm gas}$ and  $\Sigma_{\rm solids}$ are respectively the values of the gas surface density at 5.2 AU, and the product of this surface density by the dust-to-gas ratio. These two latter parameters are equivalent to the total disc mass and the dust-to-gas ratio. Using these five parameters, we can easily define a metric in the space of disc parameters by using the Euclidian distance in this 5-dimension space. In order to avoid too large differences between the scales of the different parameters, we use the logarithm of the disc parameters, and scale them in order to obtain the same mean and variance for all these quantities. This is arbitrary since it gives all the parameters the same importance in the determination of the metric, but since it is not clear which disc parameter is most important, this is a natural and conservative choice.

\begin{figure}
\center
 \includegraphics[height=0.25\textheight,angle=0]{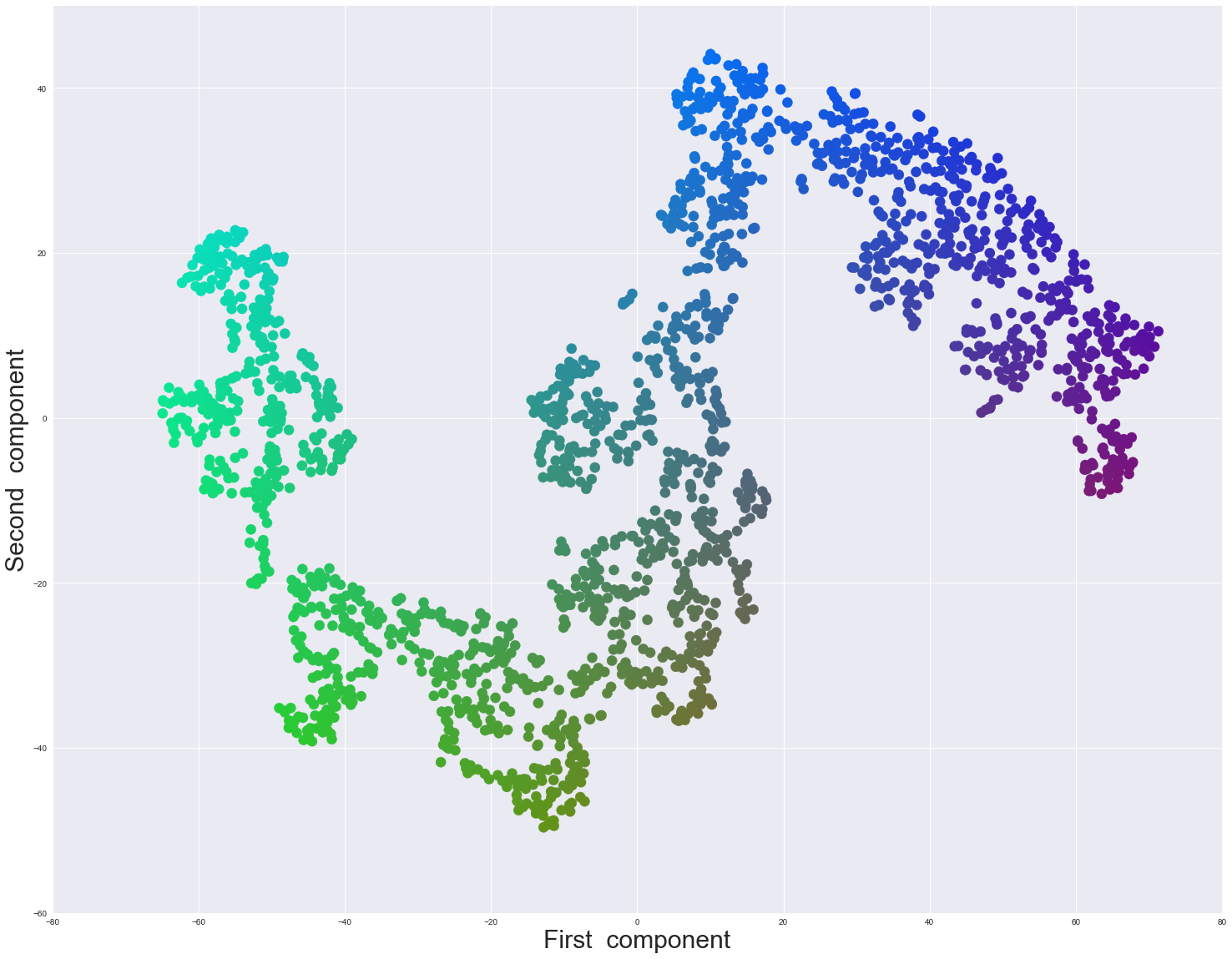}

   \includegraphics[height=0.25\textheight,angle=0]{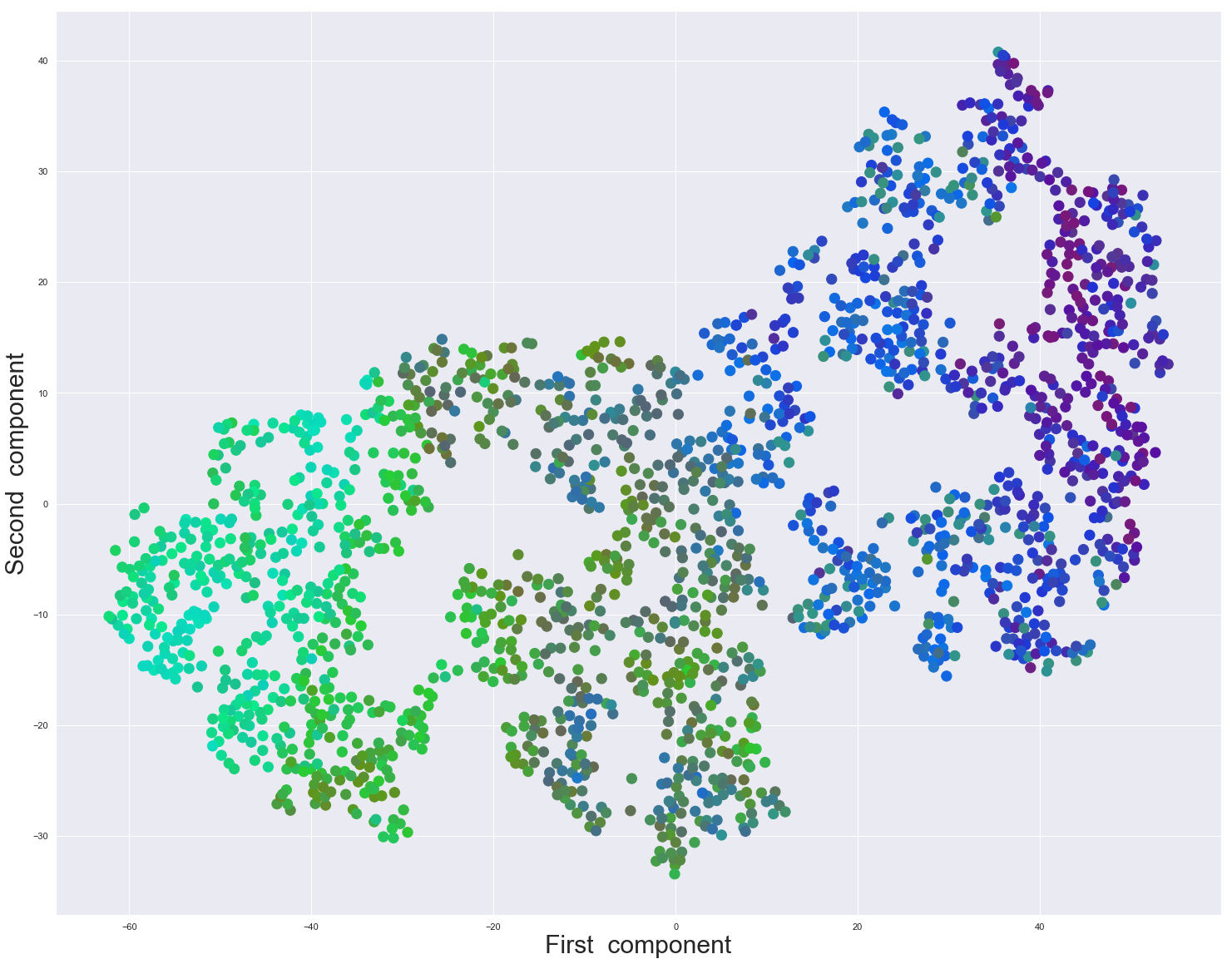}
\caption{T-SNE representation based on the distance in the space of systems (upper panel) and distance in the space of disc parameters (lower panel). The upper panel is similar to Fig. \ref{figTSNE}, upper panel, except that the colour code is here only linked to the position of the point on the plot. On the lower panel, we used T-SNE based on the similarity resulting from the metric in the space of disc parameters (see text) to represent systems. The colour code indicates in which part of the upper panel the same system is represented. In the lower panel, two points located close one to the other represents planetary systems formed in similar discs, whereas two points with similar colours represent planetary systems that are themselves similar.}
  \label{correlation}
\end{figure}

Using this new metric, we can run T-SNE to compute a 2-dimension embedding of the disc models (which originally belong to a 5-dimension space), and infer wether the embedding can be related to the one computed using the metric in the \textit{planetary system} space. For this, we have first assigned a colour to each of the points in Fig. \ref{figTSNE} which is only related to their position on the figure (see Fig. \ref{correlation}, upper panel). Then, we plot in Fig. \ref{correlation}, lower panel, the T-SNE embedding (computed using the \textit{disc parameter} metric), using the same colours. As can be seen on the figure{\footnote{An animation showing the transition from the first to the second representation can be found at \url{http://nccr-planets.ch/research/phase2/domain2/project5/machine-learning-and-advanced-statistical-analysis/}}}, the colour gradient on the upper panel is largely preserved on the lower panel. For example,  light green-blue points (which correspond to 10-planet systems with small masses) on the upper panel are preferentially located in the lower left region of the lower panel. The discs in which these low-mass 10-planets systems form are therefore similar (in term of the disc parameter metric). The colour gradient is however not totally preserved, as some local variations of the colours can be seen. This shows that other parameters (e.g. the initial location of the planetary embryos) are also important in governing the final architecture of planetary systems. 

\begin{figure}
\center
  \includegraphics[height=0.35\textheight,angle=0]{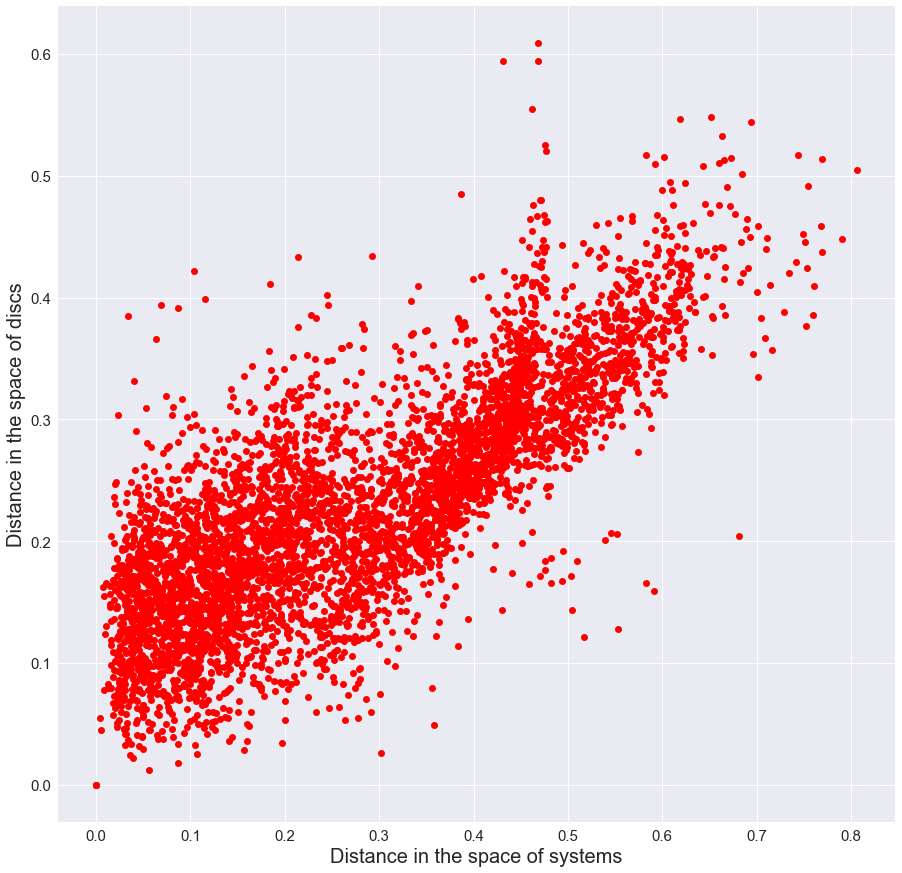}
\caption{Correlation between the inter systems distance (horizontal axis) and the distance computed using the disc parameters (vertical axis) for a subset of 5000 points.}
  \label{distances-correlation}
\end{figure}

This finding is confirmed in Fig. \ref{distances-correlation}, where we show the relation between the inter planetary system distance (using the metric introduced in Sect. \ref{subsec:weighted_distance}), and the distance between the same systems computed using the disc parameters. The correlation between both distances is clear, with some notable dispersion which is  due to the other initial conditions of the formation calculations like the starting location of the planetary embryos.

In conclusion, the disc parameters govern the trend in the planetary system architecture, while some special circumstances (e.g. special starting locations of planetary embryos) can lead to local variations.

\section{Discussion and conclusion}
\label{sec:conclusion}

\begin{figure}
  \includegraphics[height=0.35\textheight,angle=0]{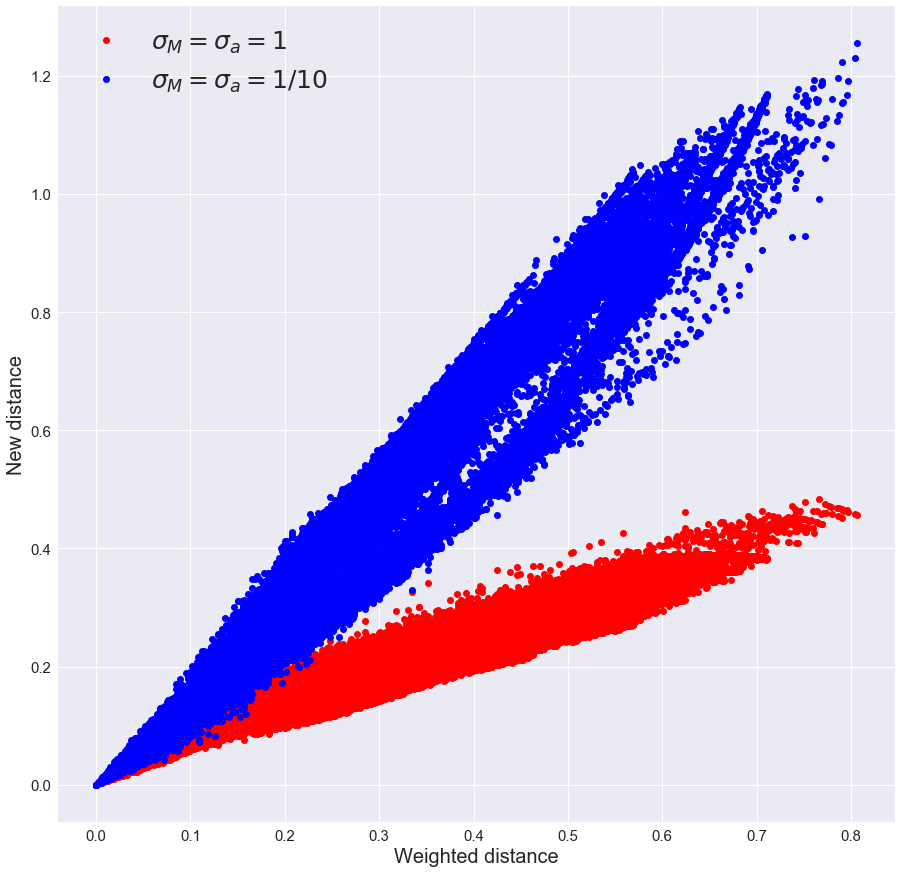}
\caption{Effect of changing the parameters in the computation of the $\Psi$ function. 5000 inter-systems distances are computed for $\sigma_M = \sigma_a = 1$ (red) and $\sigma_M = \sigma_a = 1/10$ (blue) and compared with the distance using reference values for the same systems. The correlation between the reference and the new distance is clearly seen.}
  \label{effect_param}
\end{figure}

We have presented in this paper a new metric to compare planetary systems, and have illustrated it with two synthetic populations of planets. The distance we have defined has two free parameters in the definition of the $\Psi$ functions (parameters $\sigma_M$ and $\sigma_a$). The value to be chosen for these two parameters is not very important, as the correlation between two sets of distances (computed using two set of parameters) is very strong (see Fig. \ref{effect_param}). Indeed, we emphasise the fact that it is the ranking of the distances between systems that is important (which system is at a larger distance than another to a reference one), and not the absolute value of the distance between two systems. {We also emphasis the fact that the study presented in this paper relates vectors \textit{representing} planetary systems on one side, and protoplanetary discs on the other side. Whether these vectors are accurate representations of \textit{real} planetary systems and protoplanetary discs remains to be established. }

Using the distance in the space of planetary systems, we have shown that population synthesis models produce a structure in the architecture of planetary systems that is intrinsically different from what would be obtained by just drawing at random a set of up to ten planets taken from the global population of planets. In addition, we have shown that the similarity between systems is related to the similarity between the protoplanetary discs in which they form. We have not further studied the population of systems beyond these two cases, since the population we used as an example here will be updated in a a future paper (Emsenhuber et al., in prep). The detailed study of the architecture of systems using the methods presented here will be the subject of a forthcoming paper (Alibert et al., in prep).

The metric we have presented in this paper encapsulates the comparison between masses and semi-major axis of all planets in different systems. It can be easily extended to the case where more properties are known for each planet (e.g. their radius) in a straightforward way. We note however that as the number of features increases, the number of data needed to analyse results of simulations or observations grows exponentially and can rapidly become untraceable. This is an effect of the well known 'curse of dimensionality' (e.g. Goddfellow, Bengio and Courville, 2016).

Some aspects of the architecture of planetary systems are not included in this metric. This is the case for example for the presence of resonances (e.g. mean motion resonances) in systems. Indeed, a slight variation of the semi-major axis of one planet in a system does not modify strongly the $\Psi$ function of this system. As a consequence, two systems can be very similar (according to the metric we propose in the present paper) despite the fact that one could be in mean motion resonance, and the other be out of resonance. Taking into account the resonant configurations of planetary systems requires modifications of the metric by, in essence, adding some extra dimension to the system. This will be explored in a future paper (Alibert et al., in prep).

We finally emphasise that we have weighted the contribution of planets in the determination of the $\Psi$ function using the logarithm of the mass. This is arbitrary and may not be the optimal choice. Indeed, such a choice means that the mass of a small planet in a system is not very important. On the other hand, in the precise case of the solar system, the fact that Mars has a small mass is one of the basis of our present understanding of the formation of the solar system (see Walsh et al. 2011 in the framework of the Grand Tack model). In this case, it would be legitimate to believe that our solar system, and the same system where Mars would have a mass similar to the one of the Earth could be fundamentally different (at least in term of their formation process). However, in the case of exoplanets, the level of detail of present observational constraints is far from the level of details we have for the solar system, and the metric presented in the present paper, although it will certainly have to be improved in the future, gives a useful framework to compare the global trends of models and observations.

\acknowledgements

We thank Julia Venturini for enlightening discussions.This work has been carried out within the frame of the National Centre for Competence in Research PlanetS supported by the Swiss National Science Foundation. The author acknowledges the financial support of the SNSF.

\end{document}